\documentclass[twocolumn]{openjournal}

\usepackage{physics}
\usepackage{siunitx}
\usepackage[breaklinks,colorlinks,citecolor=blue,urlcolor=blue]{hyperref}
\usepackage{natbib}

\makeatletter
\def\frontmatter@title@below{\vspace*{-2.63\baselineskip\vspace*{0.10in}}}%
\def\frontmatter@above@affilgroup{\vspace*{0.20in}}%
\makeatother

\newcommand\Alfven{Alfv\'{e}n }
\newcommand\Alfvenic{Alfv\'{e}nic }
\newcommand{\Bout}{B_0 }
\newcommand{\Bin}{B_\mathrm{in} }
\newcommand{\bin}{b_\mathrm{in} }
\newcommand{\Binout}{\flatfrac{\Bin}{\Bout}}
\newcommand{\Bbkg}{B_\mathrm{bkg} }
\newcommand{\CA}{C_\mathrm{A} }
\newcommand{\CD}{C_\mathrm{D} }
\newcommand{\dB}{\delta B }
\newcommand{\MA}{M_\mathrm{A} }
\newcommand{\phim}{\phi_\mathrm{m} }
\newcommand{\PhiL}{\Phi_\mathrm{L} }
\newcommand{\Reff}{R_\mathrm{eff}}
\newcommand{\rma}{r_\mathrm{m}}
\newcommand{\rmag}{r_\mathrm{mag}}
\newcommand{\rout}{r_\mathrm{out}}
\newcommand{\thetam}{\theta_\mathrm{m} }
\newcommand{\vA}{v_\mathrm{A} }
\newcommand{\vAn}{v_{\mathrm{A}0} }
\newcommand{\SIdeg}[1]{\SI{#1}{\degree} }
\newcommand{\degperp}[1]{#1^\circ_\perp }

\newcommand{\plane}[2]{$#1$\nobreakdashes-$#2$~plane}

\graphicspath{ {./figures/} }

\newcommand{\figsketch}{
	\begin{figure*}
	\begin{center}
	\includegraphics[width=\textwidth]{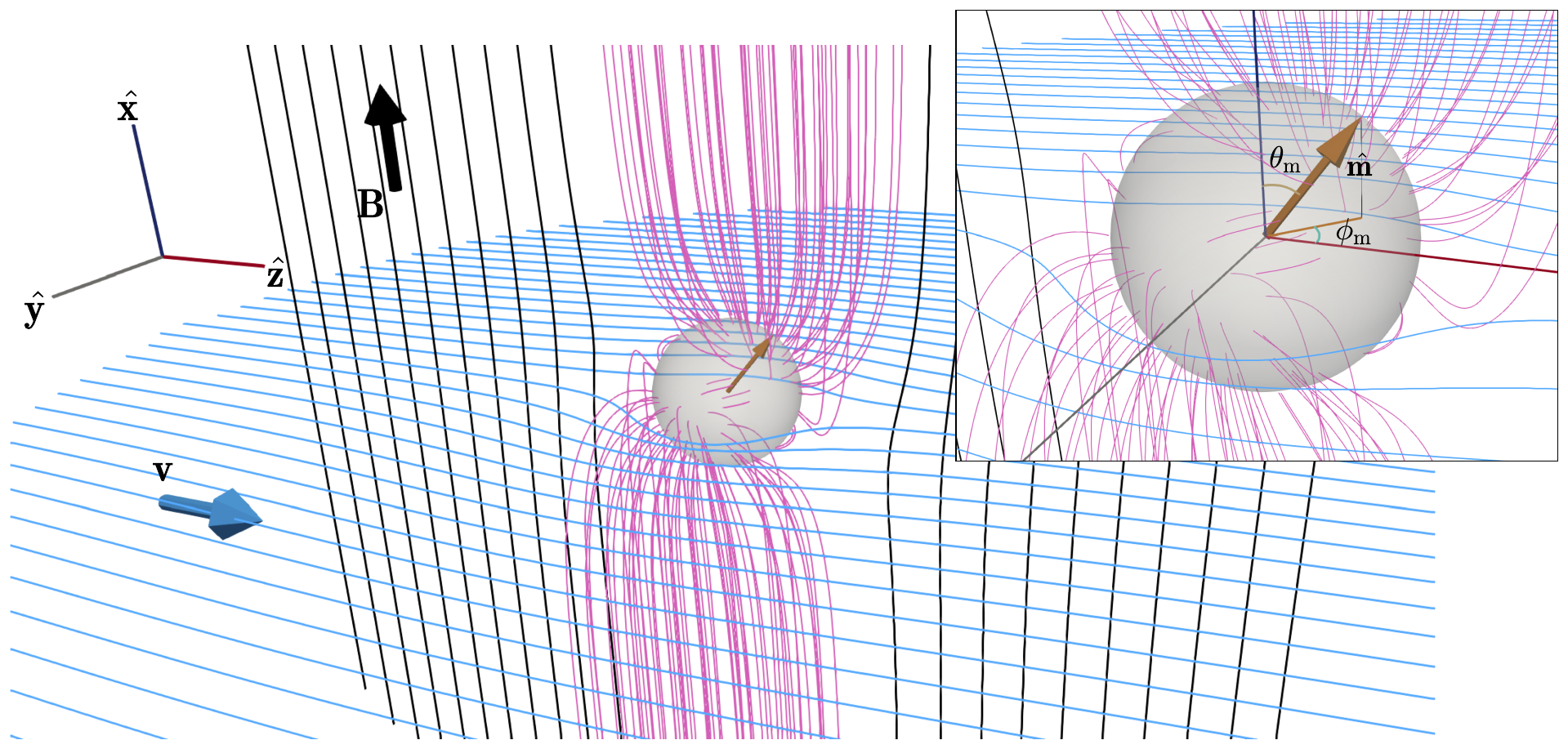}
	\caption{A sketch of our simulation setup. The gray sphere in the center
          represents the conductor of radius $R$. The black lines are open
          magnetic field lines in the \plane{x}{z}, which asymptotically
          approach $\vb{B}=B_0\vu{x}$ for $r\gg R$. The magenta lines are
          additional magnetic field lines that connect to the conductor,
          distributed with uniform flux on the conductor's surface. The blue
          lines are the velocity field lines in the \plane{y}{z}, which
          similarly approach an asymptotic value of $\vb{v}=v_0 \vu{z}$ for
          $r\gg R$. The inset zooms in on the conductor, showing the parameters
          of the dipole moment's orientation, $\vu{m}$, indicated by an orange
          arrow. The polar angle, $\thetam$, is measured from $\vu{x}$, and the
          azimuthal angle, $\phim$, is measured in the \plane{y}{z} from
          $\vu{z}$.}
			\label{fig:sketch}
		\end{center}
	\end{figure*}
}

\newcommand{\figfidvisual}{
	\begin{figure*}
	\begin{center}
	  \includegraphics[width=0.9\textwidth]{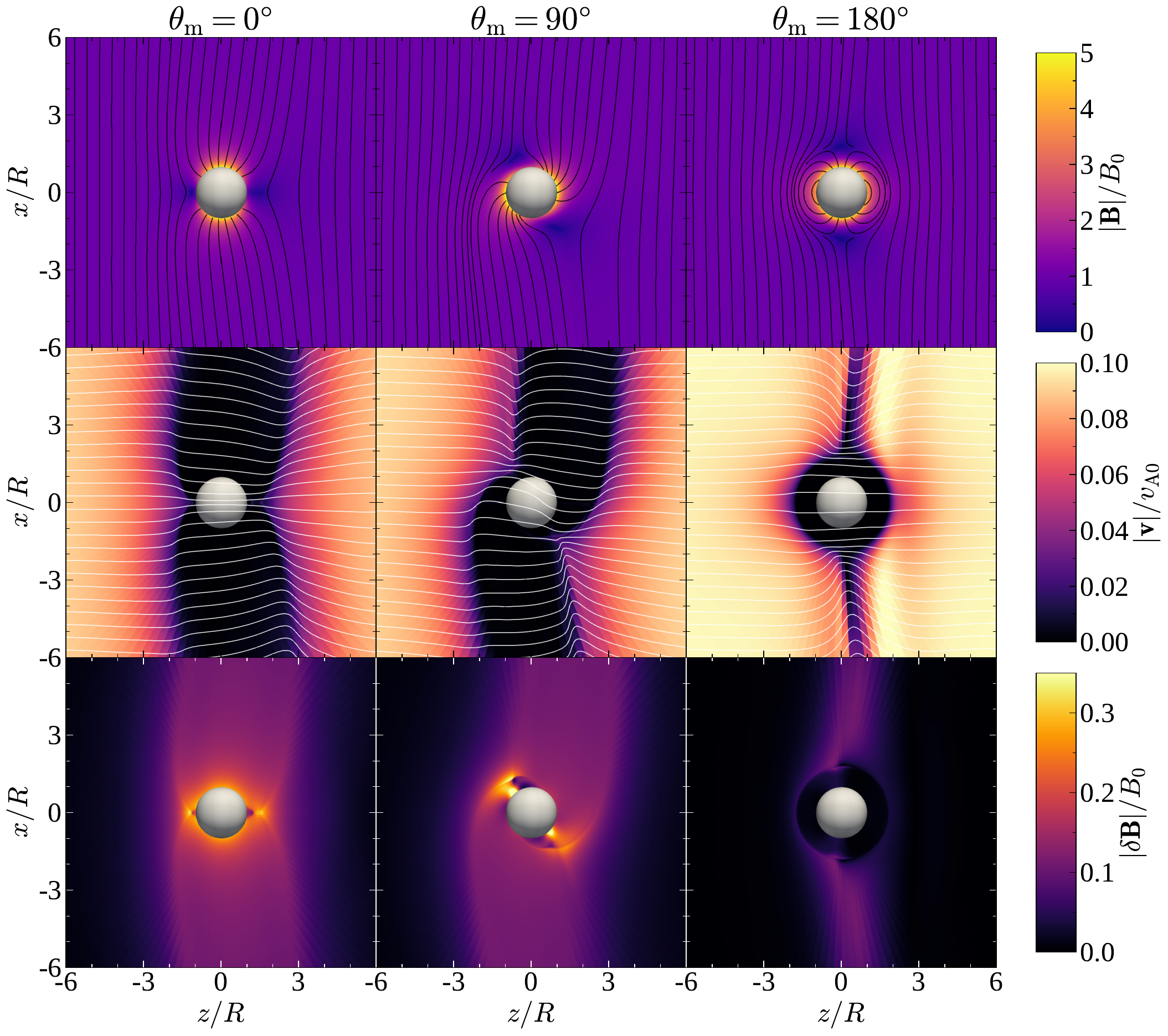}
	\caption{Diagnostic quantities for our fiducial runs, with $\MA=0.1$,
          $\beta=2$, $\bin=5$, on the \plane{x}{z} after the simulation has
          achieved a steady state, for three angular configurations:
          $\SIdeg{0},\SIdeg{90},\SIdeg{180}$ (left, middle, and right columns,
          respectively). The quantities are $\vb{B}$ (top row), $\vb{v}$ (middle
          row), and $|\vb{\delta B}|$ (bottom row). The conductor is shown as a
          gray sphere in the center of the domain. Vector quantities are plotted
          with field lines in a colored background which displays magnitude.
          Each field line begins at $r=10R$ and $y=0$. If a field line travels
          too far out-of-plane (away from $y=0$), it is truncated for
          visibility. All evolved quantities show disturbances along the \Alfven
          wing characteristics. We note that the velocity field lines wrap
          around the wings since motion stops entirely along the
          characteristics.}
			\label{fig:fidvisual}
		\end{center}
	\end{figure*}
}

\newcommand{\figCAB}{
\begin{figure}
\begin{center}
  \includegraphics[width=\columnwidth]{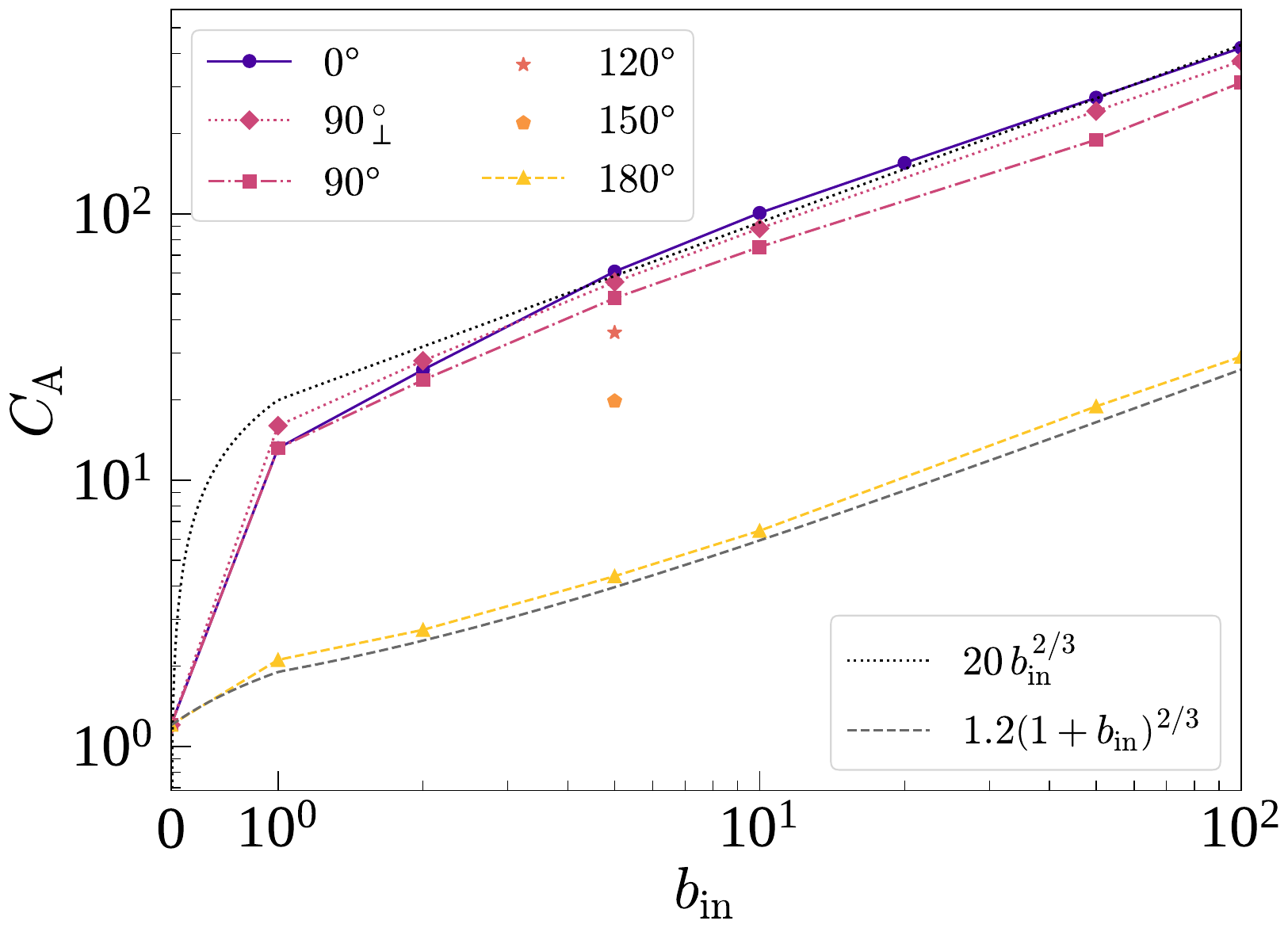}
\caption{The drag coefficient $\CA$ as a function of $\bin$ for a range of
  inclination angles, with fixed $\MA=0.1$ and $\beta=2$. We run a full suite of
  simulations for the angular configurations
  $\theta_m=\SIdeg{0},\,\SIdeg{90},\,\SIdeg{180}$, all with $\phi_m=\SIdeg{0}$,
  and $\theta_m=\SIdeg{90},\,\phi_m=\SIdeg{90}$ (labeled by $\degperp{90}$). We
  also show the results for $\theta_m=\SIdeg{120}, \SIdeg{150}$ and $\bin=5$.
  The gray dashed curve shows our proposal for the scaling in the anti-aligned
  system (Eq.~\ref{eq:antidrag}), and the black dotted line shows an approximate
  fit to the aligned case, given by Eq.~(\ref{eq:aldrag}).}
			\label{fig:CAB}
		\end{center}
	\end{figure}
}

\newcommand{\figanglecomp}{
\begin{figure}
\begin{center}
\includegraphics[width=\columnwidth]{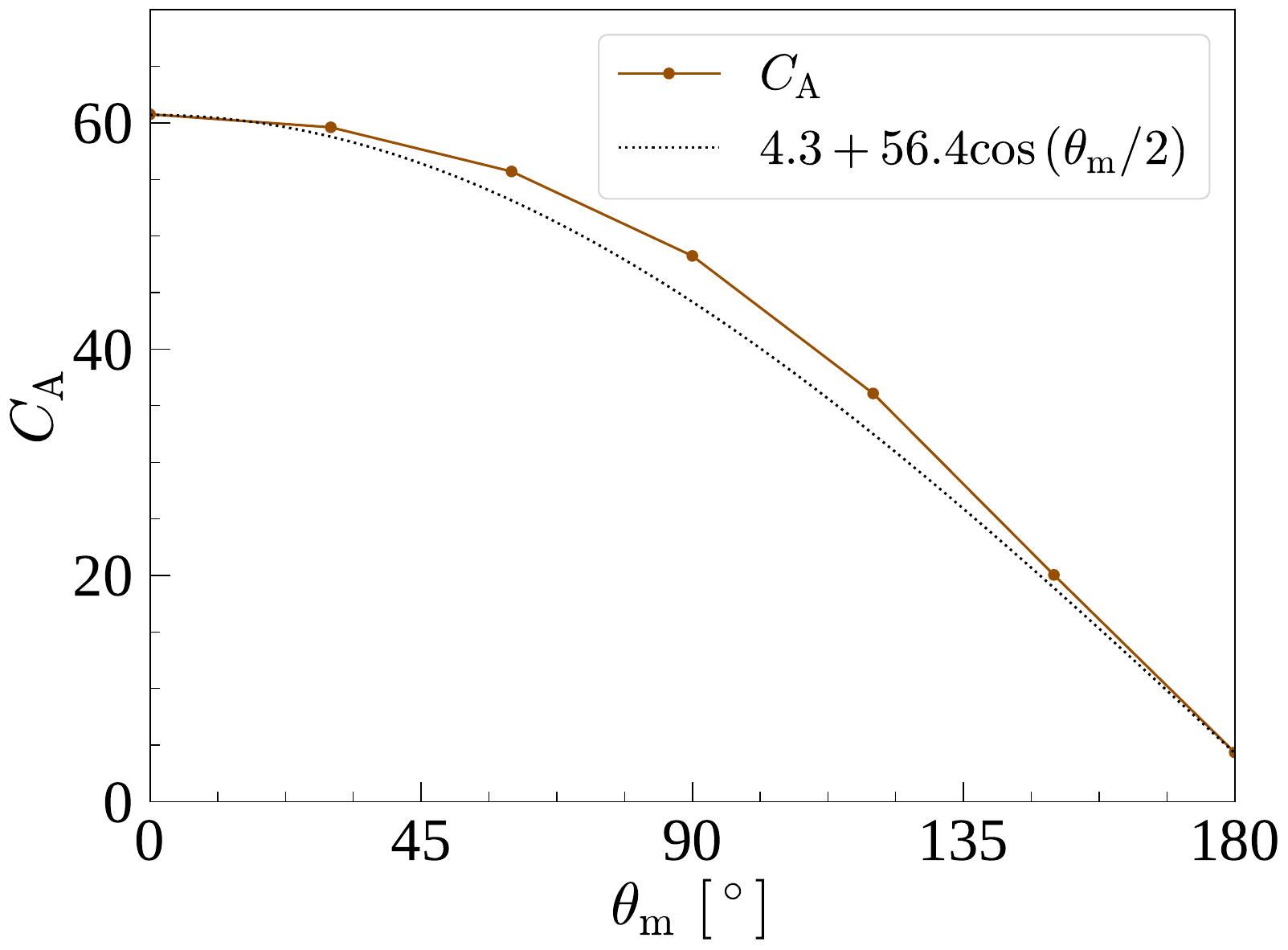}
\caption{The drag coefficient $\CA$ as a function of $\thetam$, for fixed
  $\bin=5$, $\MA=0.1$, $\beta=2$, and $\phim=\SIdeg{0}$. The black dotted line
  shows an approximate fit to the data.}
			\label{fig:anglecomp}
		\end{center}
	\end{figure}
}

\newcommand{\figvcomp}{
\begin{figure}
\begin{center}
\includegraphics[width=\columnwidth]{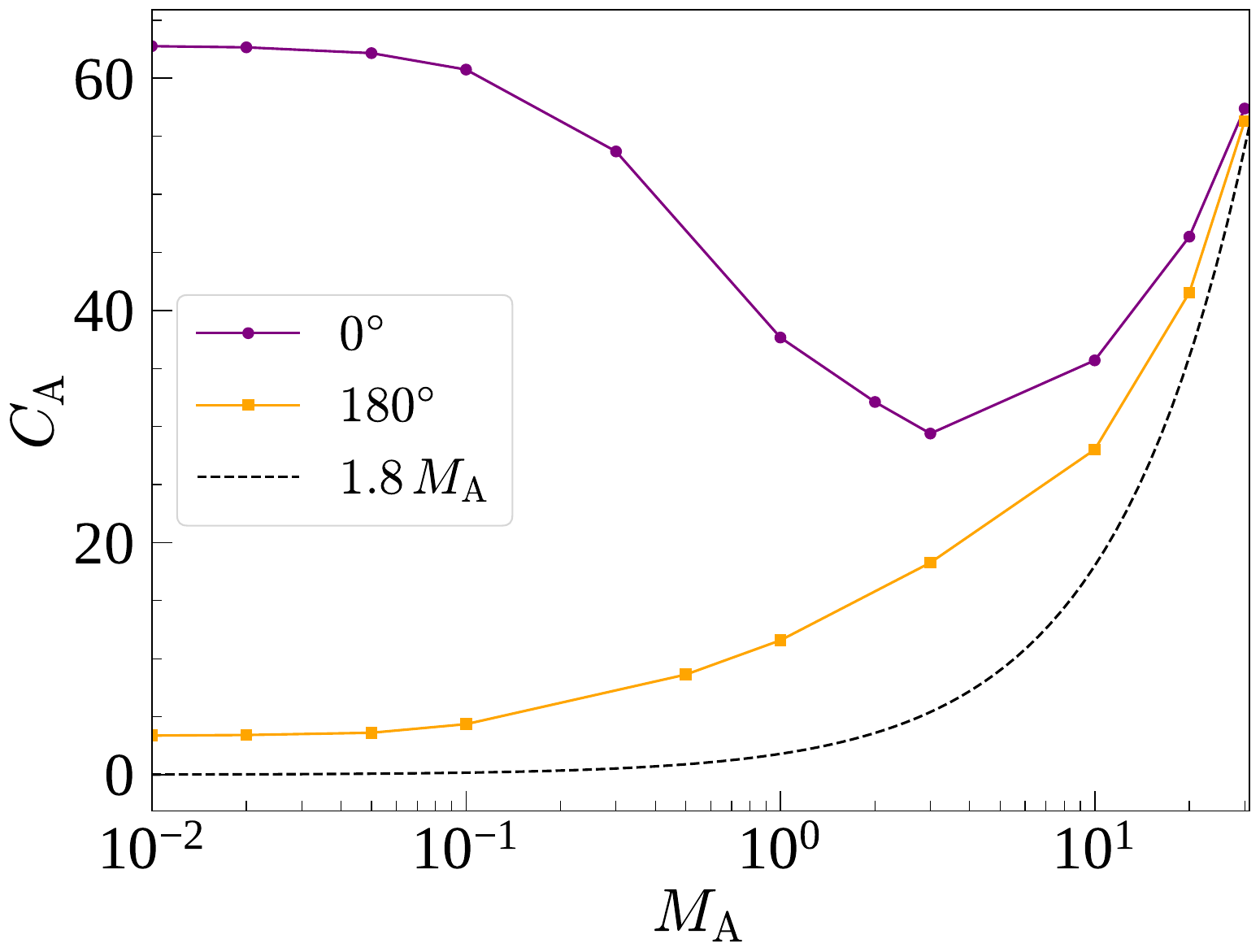}
\caption{The drag coefficient $\CA$ as a function of $\MA$ for fixed $\bin=5$
  and $\beta=2$ and for both the aligned (top, purple) and anti-aligned (bottom,
  yellow) cases. The black dashed curve shows our estimate for the asymptotic
  scaling of $\CA\propto\MA$.}
			\label{fig:vcomp}
		\end{center}
	\end{figure}
}

\newcommand{\figfieldlinesurvey}{
\begin{figure*}
\begin{center}
\includegraphics[width=0.8\textwidth]{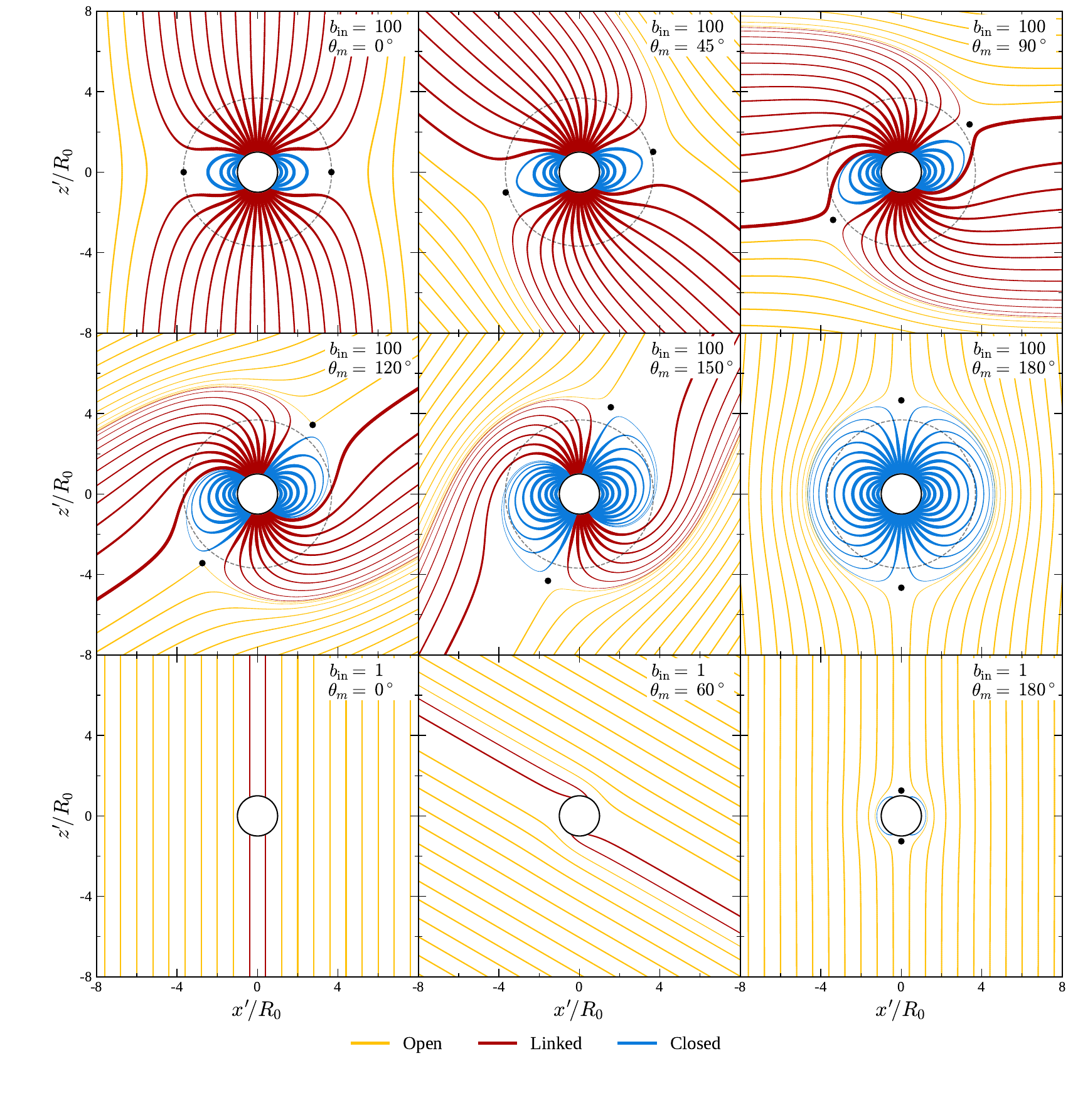}
\caption{Field line configuration of $\vb{\Bbkg}$ for a variety of dipole
  orientations and field strengths. The top two rows have $\bin=100$, with
  $\thetam=\SIdeg{0},\SIdeg{45},\SIdeg{90},\SIdeg{120},\SIdeg{150},\SIdeg{180}$
  (left to right, top to bottom). The bottom row has $\bin=1$ and
  $\thetam=\SIdeg{0},\SIdeg{60},\SIdeg{180}$. We color the fully open field
  lines yellow, the open field lines ``linked'' to the conductor red, and the
  closed field lines blue. All plots are rotated from the original coordinate
  system to a new one such that the $z^\prime$-axis is aligned with the dipole
  moment and ${\bf B}_0$ lies in the \plane{x^\prime}{z^\prime}. All field lines
  remain in-plane, but we vary their thicknesses to correspond to changes in
  field line density in the $y^\prime$ direction. A circle of radius $\rma$ (see
  Eq.~\ref{eq:rm}) is shown in each plot for a sense of scale, and the magnetic
  null points are marked with black dots.}
		\label{fig:fieldlinesurvey}
		\end{center}
	\end{figure*}
}

\newcommand{\figfieldlineflux}{
\begin{figure*}
  \begin{center} 
\includegraphics[width=\columnwidth]{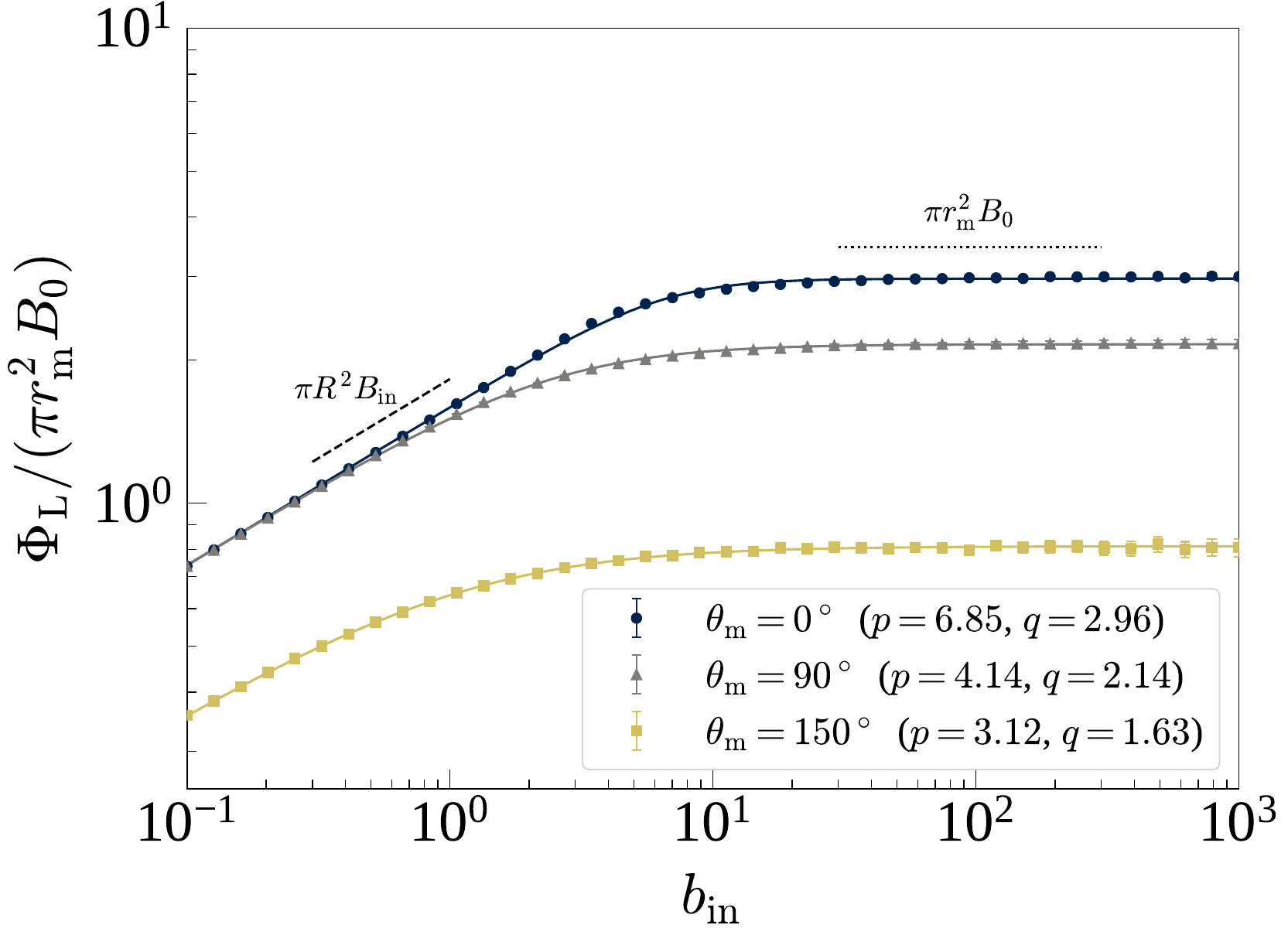}\hfill%
\includegraphics[width=0.975\columnwidth]{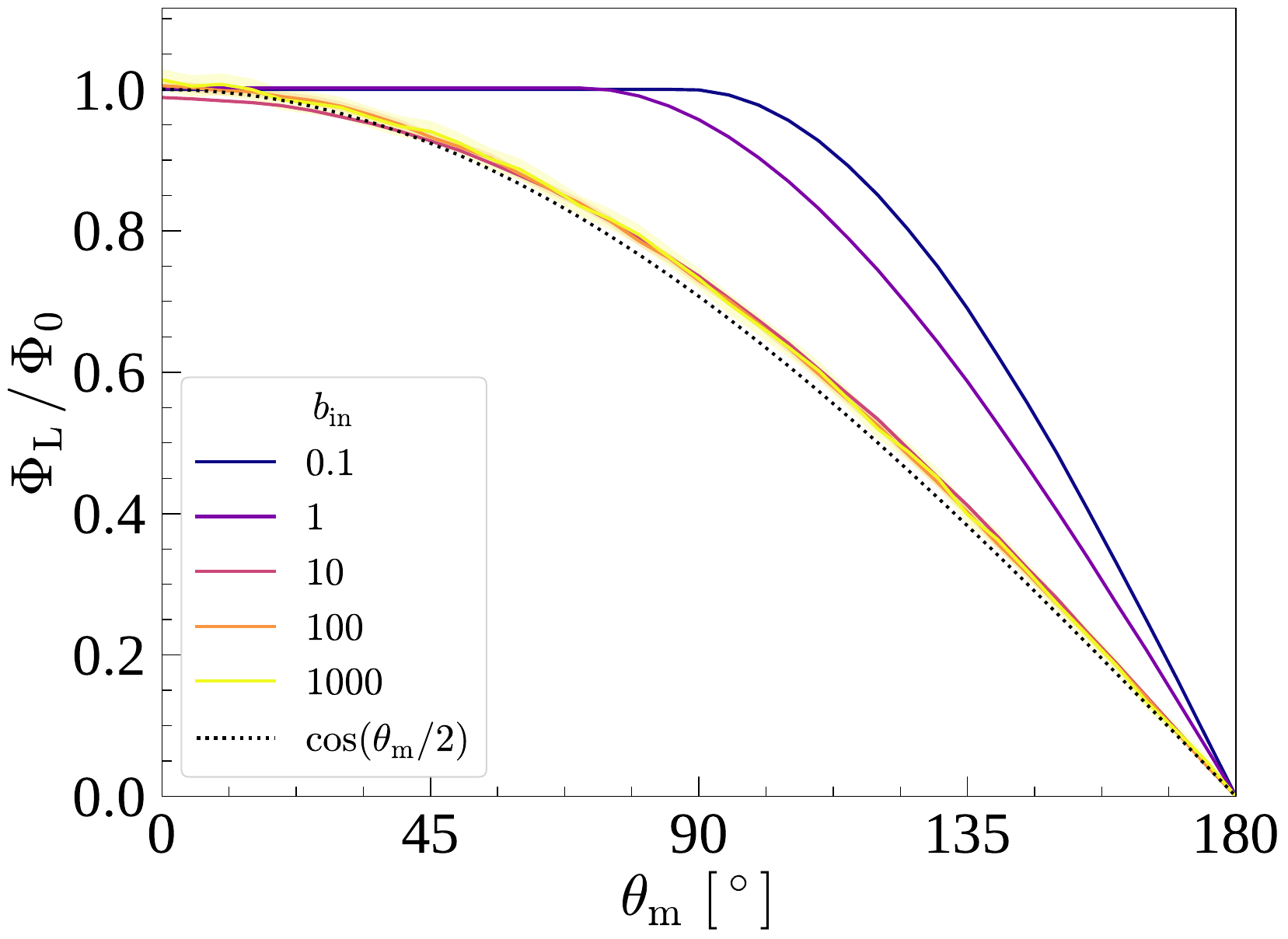}
\caption{\textit{Left}: The flux of open $\Bbkg$ field lines linked to the
  conductor, $\PhiL$, as a function of $\bin$, for
  $\thetam\in\{\SIdeg{0},\SIdeg{90},\SIdeg{150}\}$. Error bars stem from the
  error of our numerical calculation of the field line flux. The data for a
  given $\thetam$ is fit according to Eq.~(\ref{eq:phi0ap}), with the values of
  the fitting parameters $p$ and $q$ given in the legend. The black dashed and
  dotted lines visualize the limiting scaling laws for $\PhiL$. \textit{Right}:
  $\PhiL$ as a function of $\thetam$, for $\bin\in[0.1,1000]$. The linked flux
  is normalized by $\Phi_0$, the value for the aligned case, such that all
  curves collapse to a single point at $\thetam=\SIdeg{0}$. A fitting curve
  following $\cos(\thetam/2)$ is depicted with a black dotted line.}
			\label{fig:fieldlineflux}
		\end{center}
	\end{figure*}
}

\newcommand{\figfidv}{
	\begin{figure*}
		\begin{center}
			\includegraphics[width=\textwidth]{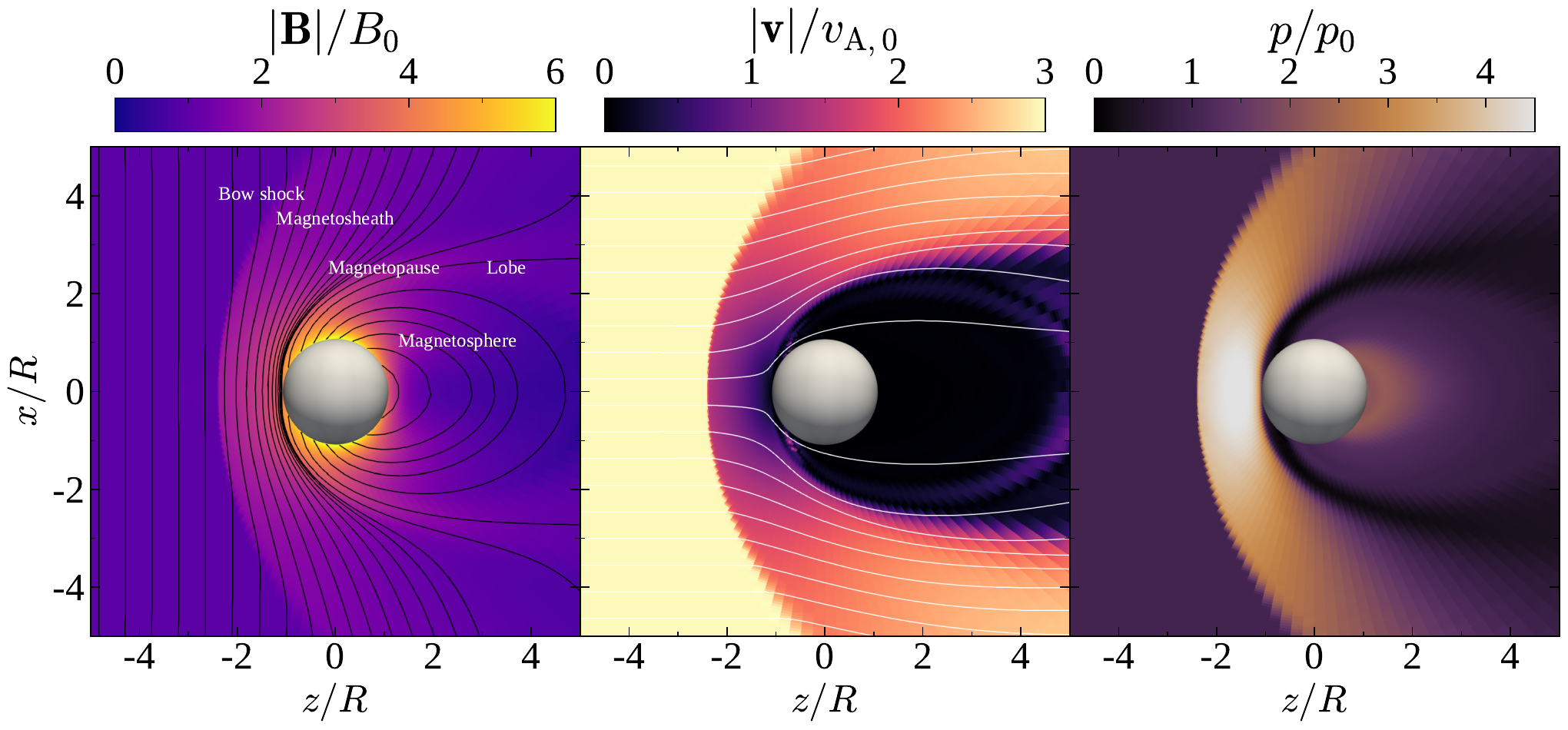}
			\caption{Diagnostic parameters for our fiducial super-\Alfvenic run,
				with $\bin=5$ (aligned), $\MA=3$, and $\beta=2$, on the
				\plane{x}{z} after the simulation has achieved a steady state.
				The quantities are $\vb{B}$ (left), $\vb{v}$ (middle), and $p$
				(right). We have labeled the major features of this system in
				the left panel. Note that the magnetotail is too far downstream
				to be visible in the panel. The conductor is shown as a gray
				sphere in the center of the domain. Vector quantities are
				plotted with field lines placed in front of a background which
				displays magnitude. Each field line begins at $r=10R$ and $y=0$.
				In the middle panel, velocity flows out of the plane (away from
				$y=0$) to divert around the magnetopause.}
			\label{fig:fidv}
		\end{center}
	\end{figure*}
}

\newcommand{\orcidauthor}[3]{\author{\href{https://orcid.org/#1}{#2$^{#3}$}}}

\journalinfo{The Open Journal of Astrophysics}
\shorttitle{Simulations of a Conducting Sphere in Magnetized Plasma II}
\shortauthors{Corso \& Lai}

\begin{document}

\title{Simulations of a Conducting Sphere Moving through Magnetized Plasma II:
Effect of Internal Magnetic Fields and Enhancement of Alfv\'{e}nic Drag}

\orcidauthor{0000-0002-0088-2563}{Nicholas J. Corso}{1*}
\orcidauthor{0000-0002-1934-6250}{Dong Lai}{2,1\dagger}
\affiliation{$^{1}$Department of Astronomy, Center for Astrophysics and
Planetary Science, Cornell University, Ithaca, NY 14853, USA\\
$^{2}$Tsung-Dao Lee Institute, Shanghai Jiao-Tong University, Shanghai, China}

\thanks{$^*$E-mail: \href{mailto:njc86@cornell.edu}{njc86@cornell.edu};
$^\dagger$E-mail: \href{mailto:dl57@cornell.edu}{dl57@cornell.edu}}

\begin{abstract}
Magnetic interaction between astrophysical objects can play an important role
and produce electromagnetic radiation in various systems, ranging from
planet-moon and star-planet systems to binary compact objects. We extend our
previous numerical study of an ideal magnetized plasma flowing past a conducting
sphere to the case where the sphere is magnetized. Such a flow generates
magnetic disturbances and produces a drag force on the sphere. We examine the
drag coefficient as a function of the strength and orientation of the
conductor's internal magnetic field relative to the background magnetic field.
We find that a strong internal magnetic field that is aligned with the
background field significantly enhances the drag coefficient compared to the
unmagnetized case; this enhancement is related to the increased magnetic flux
that connects the conductor and the open, external field region. We also explore
how the drag coefficient changes as the flow transitions from the sub-\Alfvenic
to the super-\Alfvenic regime. Our results suggest that magnetic interaction may
play a larger role in orbital evolution of some binary systems than previously
expected.
\end{abstract}


\section{Introduction} \label{sec:intro}

Since the discovery of radio emission and auroral signatures caused by Io
cutting through Jupiter's magnetosphere \citep{Bigg64} and their first
theoretical study \citep{Goldreich&Lynden-Bell69}, there has been considerable
interest in observing similar interactions between magnetized astrophysical
objects in binaries. These include planet-moon systems
\citep{Neubauer90,Bagenal&Dols20}, star-planet systems \citep{Li+98,Preusse+06,
  Lanza09,Laine&Lin11,Lai12,Lanza13,Strugarek+14,Saur18,DeColle+25,Lee&Owen25}
and binary compact objects \citep{Hansen&Lyutikov01,Lai12,Most&Philippov22,
  Most&Philippov23,Mahlmann&Beloborodov25}. Such interactions may result in
observable electromagnetic radiation and/or lead to long-term orbital evolution.

A related problem concerns an unmagnetized or weakly magnetized object moving in
the more dominant magnetic field of its companion. In this case, the weaker
companion can be treated as a perturber of the global background field, allowing
for a clean local analysis of this object and its environment. Such an analysis
began with \cite{Drell+65}, who considered the model of a conductor (treated as
a rectangular bar) moving with velocity $\vb{v_0}$ through a plasma with a
perpendicular uniform magnetic field $\vb{B_0}$. Because of the sub-\Alfvenic
velocity ($v_0<\vA$, where $\vA$ is the \Alfven speed of the background plasma),
the conductor acts as an electrical generator that launches two tube-like
\Alfven waves---termed \Alfven wings---that propagate outward along the
background field lines, carrying power that drags the conductor. The concept of
``\Alfven wings'' has found application in the Jupiter-Io system
\citep{Neubauer80}, where \Alfven waves could propagate back-and-forth between
the two objects as though in a closed circuit. More recently, it has been
applied more generically, such as in exoplanetary systems (\textit{e.g.}
\citealt{Neubauer98,Strugarek+14,DeColle+25}).

Many previous studies have examined the detailed dissipative mechanisms and
radiative signatures of magnetic interactions. In planet-moon systems, such as
Jupiter-Io, the power carried by \Alfven wings can produce aurorae and radio
emissions \citep{Neubauer90,Bagenal&Dols20}. In star-planet systems, ohmic
dissipation from the current carried by \Alfven wings may drive coronal activity
or heating in either body \citep{Li+98,Preusse+06,Lanza09,Laine&Lin11}, and the
flux tube connecting the two bodies may reconnect and flare out
\citep{Lai12,Lanza13}. Similar processes may also operate in compact-object
binaries \citep{Hansen&Lyutikov01,Lai12,Wang+16,Beloborodov21,Most&Philippov22,
  Most&Philippov23}. Recently, \cite{Sharma+26} proposed that \Alfven wings in
merging neutron star binaries could lead to pulsar-like emission due to strong
currents at the wing boundaries; \cite{Lyutikov26} followed up this proposal
with kinetic simulation. These systems may also experience interactions not
mediated by \Alfven wings. For instance, \cite{Mahlmann&Beloborodov25} suggested
that in a binary neutron star system with aligned spins, turbulent heating could
arise from the Kelvin--Helmholtz instability at the interface of their
magnetospheres.

\figsketch

This paper is the second in a series, following \cite{Corso&Lai26} (hereafter
Paper I), where we study numerically the interaction between a moving conductor
and an (otherwise homogeneous) magnetized plasma. We consider an ideal MHD model
which is scale-independent, even if it may not capture the nuances of specific
binary systems highlighted above. In Paper I, we focused on the case of an
unmagnetized (i.e. zero internal magnetic field) conductor [see also
\cite{DeColle+25,Sharma+26}], and calculated the drag force $F$ on the
conductor. We introduced the dimensionless drag coefficient, $\CA$, such that
the mechanical power, $P=Fv_0$, imparted by the flow on the conductor is given
by
\begin{equation}
    P=\CA\rho_0 R^2 v_0^2 \vAn, \label{eq:drag}
\end{equation}
where $\rho_0$, $v_0$, and $\vAn=(B_0^2/4\pi\rho_0)^{1/2}$ are the background
values for the density, velocity, and \Alfven speed, and $R$ is the radius of
the conducting sphere. Using MHD simulations, we determined how $\CA$ depends on
the \Alfven Mach number and the plasma $\beta$ parameter:
\begin{equation}
    \MA=\frac{v_0}{\vAn},\quad \beta=\frac{p_0}{(B_0^2 / 8\pi)},
\end{equation}
where $p_0$ and $B_0$ are the background values for gas pressure and the
magnetic field. We validated the basic picture of the \Alfven wing model by
demonstrating that $\CA$ is approximately constant for $\MA \lesssim 1$. We also
showed that slow magnetosonic waves provide a correction to the picture through
their own wing-like features.

In this paper, we extend our previous investigation to the case where the
conductor has an internal magnetic field $\vb{\Bin}$. Not only does this extend
the range of applicability of our model to include the cases where the companion
has its own magnetosphere, but we will show that the internal field can
substantially increase the drag imparted on the conductor. As in Paper I, our
primary goal is to obtain the general scaling relation for the drag coefficient
that can be applied to different astrophysical and planetary contexts.

This paper is organized as follows: in Section \ref{sec:method}, we review the
numerical method for our simulations and introduce the problem setup; in Section
\ref{sec:subalfvenic}, we present our results for the sub-\Alfvenic regime and
model the power dissipation using a semi-analytical treatment; in Section
\ref{sec:supalfvenic}, we present our results for the super-\Alfvenic regime; we
conclude with Section \ref{sec:discuss}.

%%%%%%%%%%%%%%%%%%%%%%%%%%%%%%%%%%
\section{Problem Setup and Method} \label{sec:method}

Our problem setup is illustrated in Figure \ref{fig:sketch}. We consider a
system composed of a conducting sphere of radius $R$ embedded in a plasma that
is flowing and magnetized. Far from the conductor, the magnetic field approaches
$\vb{\Bout}=\Bout\vu{x}$, and the plasma velocity approaches $\vb{v}_0=v_0
\vu{z}$. For the present study, we add an intrinsic uniform magnetic field,
$\vb{\Bin}=\Bin\vu{m}$ (where $\vu{m}$ is the unit vector), to the conductor. We
simulate our system using the magnetohydrodynamic (MHD) code, \texttt{PLUTO}
\citep{Mignone07}. We close the MHD equations with an ideal equation of state
with adiabatic index $\gamma=\flatfrac{5}{3}$.

As in Paper I, we write the magnetic field as
\begin{equation}
	\vb{B}(\vb{r},t) = \vb{\Bbkg}(\vb{r}) + \vb{\dB}(\vb{r},t),
\end{equation}
and evolve $\vb{\dB}(\vb{r},t)$ in our MHD simulation. The time-independent
background field is given by
\begin{equation}
	\vb{\Bbkg}=\Bout\vu{x}+\frac{1}{2}\left( \frac{R}{r}\right)^3
		\Bigl[\Bin \vb{d}(\vu{m}) - \Bout \vb{d}(\vu{x})\Bigr], \label{eq:Bbkg}
\end{equation}
where $\vb{d}(\vu{n})\equiv 3(\vu{r}\vdot\vu{n})\vu{r}-\vu{n}$ is a vector
representing a dipole field configuration with a magnetic moment aligned with
$\vu{n}$. This background field satisfies the asymptotic condition
$\vb{\Bbkg}\rightarrow \vb{\Bout}$ at $r\rightarrow\infty$ and the boundary
condition at $r=R$, i.e. the radial component of the field is continuous across
the conductor's surface. We see in Eq.~(\ref{eq:Bbkg}) that $\vb{\Bin}$
introduces a dipole contribution to the external field. We express the
orientation of $\vu{m}$ relative to $\vu{x}$ (the direction of $\vb{\Bout}$)
using the polar angles $\thetam$ and $\phim$, the standard spherical coordinate
angles with respect to the $x$-axis pole, i.e.
\begin{equation}
	\vu{m} = 
	\begin{pmatrix}
		\cos\thetam,&-\sin\thetam\sin\phim,&\sin\thetam\cos\phim
	\end{pmatrix}. \label{eq:moment}
\end{equation}

\figfidvisual

With this general setup, the MHD flow outside the conductor and the associated
drag coefficient $\CA$ (defined in Eq.~\ref{eq:drag}) depend on $\MA$, $\beta$,
and $\vb{\bin}\equiv\vb{\Bin}/\Bout$, i.e.,
\begin{equation}
  \CA=\CA(\MA,\beta,\bin,\thetam,\phim). \label{eq:dragparam}
\end{equation}
As detailed in Paper I, we compute the drag by evaluating total momentum flux
across a spherical surface that encloses the conductor, and making sure that the
result is independent of the radius of the surface.

We introduce a non-slip boundary condition, demanding $\vb{v}=0$, at the
conductor's surface ($r=R$). Note that this stands in contrast to the case where
the conductor is unmagnetized, where velocity slippage is allowed. When
$\Bin\neq 0$, the normal component of the magnetic field (which must be
continuous across the surface) is nonzero. The non-slip condition is required to
ensure that the tangential component of $\vb{E}$ vanishes at $r=R$ as a
consequence of the induction equation,
\begin{equation}
	\vb{E}=-\vb{v}\cross\vb{B}. \label{induc}
\end{equation}
From Faraday's Law, this fixes the component of $\vb{B}$ normal to the surface,
maintaining a frozen flux from the conductor itself. Thus, the fixed magnetic
flux at the conductor's surface creates a boundary layer akin to that caused by
a finite viscosity. To accommodate this new boundary condition, we initialize
the velocity field as follows:
\begin{multline}
	\vb{v}=v_0\cos\theta\left[1-\frac{3}{2}\left(\frac{R}{r}\right)
		+\frac{1}{2}\left(\frac{R}{r}\right)^3\right]\vu{r} \\
	- v_0\sin\theta\left[1-\frac{3}{4}\left(\frac{R}{r}\right)
		-\frac{1}{4}\left(\frac{R}{r}\right)^3\right]\vu*{\theta},
\label{eq:vinit}\end{multline}
such that $\vb{v}$ vanishes at $r=R$ and approaches $\vb{v_0}=v_0\vu{z}$ for
$r\gg R$. Note that in Eq.~(\ref{eq:vinit}), the spherical coordinates
$(r,\theta,\phi)$ are defined with respect to the $z$-axis pole.

All simulations are run on a discretized grid using spherical coordinates
$(r,\theta,\phi)$. The inner boundary of the computational domain is set at
$r=R$, and the outer boundary at $\rout\gg R$. We use a zero-gradient outflow
condition at $r=\rout$. We also include a damping zone where evolved variables
are driven to their background values close to the outer boundary. For most
simulations, the computational domain spans $r\in[R,10R]$ and the full unit
sphere, using $N_r\times N_\theta\times N_\phi=200\times64\times128$ grid
points, where the radial points are spaced logarithmically, such that the cell
thickness spans $\Delta r\in[0.012R,0.12R]$, and the angular points are spaced
uniformly. The damping zone spans $r\in[8R,10R]$. For cases where $\bin>10$, we
extend the domain to $\rout=50R$, scaling the damping zone to $r\in[40R,50R]$
and $N_r=340$ (approximately preserving the value of $\Delta r$ close to the
conductor's surface). This extension is necessary to keep $\rout$ sufficiently
far from the effective radius of the conductor, which scales with $\bin$. See
Paper I for more details on our numerical methods.

%%%%%%%%%%%%%%%%%%%%%%%%%%%%%%%%
\section{Sub-Alfv\'enic Regime} \label{sec:subalfvenic}

In this section, we examine the sub-\Alfvenic regime ($\MA\lesssim 1$), where
\Alfven wings dominate the dynamics. This is the main focus of this paper and of
Paper I. The transition to the super-\Alfvenic regime will be discussed in
Section \ref{sec:supalfvenic}.

\subsection{Numerical Results} \label{sec:subresults}

We carry out two suites of simulations. In the first, we examine the effects of
modifying the strength and orientation of the dipole. In these simulations, we
fix $\MA=0.1$ and $\beta=2$, and we consider the internal magnetic field in the
range $\bin\in[0,100]$ and angular configurations:
$\thetam=\SIdeg{0},\SIdeg{90}, \SIdeg{180}$, all with $\phim=0$; and
$\thetam=\SIdeg{90}$, $\phim=\SIdeg{90}$ (labeled $\thetam=\degperp{90}$). We
further test the cases of
$\thetam=\SIdeg{30},\SIdeg{60},\SIdeg{120},\SIdeg{150}$ for $\bin=5$. Note that
we refer to $\thetam=\SIdeg{0}$ as the ``aligned'' case with respect to
$\vb{\Bout}$. In the binary context where the conductor is embedded in the
dominant magnetic field of a primary object, this corresponds to anti-alignment
between dipole moments of the two binary components. Our second suite of
simulations examines the effect of varying $\MA$. We consider both the aligned
and anti-aligned cases, holding $\bin=5$ and $\beta=2$, and we test values in
the range $\MA\in[0.01,30]$. Each simulation is run until $\CA$ reaches a
constant (time-independent) value, indicating that the system has reached a
steady state.

Figure \ref{fig:fidvisual} depicts key diagnostic quantities for our fiducial
runs with $\bin=5$, and $\thetam=\SIdeg{0}$, $\SIdeg{90}$, and $\SIdeg{180}$. We
notice a stark contrast between the magnetic and velocity perturbations in the
aligned and anti-aligned cases. In the anti-aligned case, where external field
lines fully wrap around the conductor's magnetosphere (where its magnetic field
dominates over $\vb{\Bout}$), the profiles of $\vb{v}$ and $\left|\dB\right|$
strongly resemble that of the unmagnetized case studied in Paper I, except that,
with $\bin=5$ the conductor has a larger effective size due to its
magnetosphere. By contrast, these profiles for the $\thetam=0,\,90^\circ$ cases
exhibit much larger disturbances. Comparing the profiles for $\vb{B}$, we can
see that the \Alfven wings trace the open field lines that connect to the
conductor.

We show the main result of our parameter survey in Fig.~\ref{fig:CAB} by
plotting $\CA$ as a function of $\bin$ for all considered $\vb{\Bin}$
orientations. As shown in Fig.~\ref{fig:fidvisual}, for the anti-aligned case,
the main effect of finite $\bin$ is the increased effective size of the
conductor due to its magnetosphere. In general, the location of the magnetopause
is determined by balancing the momentum flux $(p+\rho
v_\perp^2+B_\parallel^2/8\pi)$ across the contact surface, where $p$ is the gas
pressure, $v_\perp$ is the gas velocity perpendicular to the surface, and
$B_\parallel$ is the magnetic field parallel to the surface (e.g.,
\citealt{Shu92}). Neglecting the plasma density and pressure within the
magnetosphere, the magnetopause has a radius given by
\begin{equation}
  \rmag\simeq R\left(\frac{\bin}{2}\right)^{1/3}
    \left(1+2\MA^2+\beta\right)^{-1/6},
  \label{eq:rmag}
\end{equation}
which, for our parameter regime ($\MA\lesssim 1$, $\beta\lesssim 1$), reduces to
\begin{equation}
	r_{\rm mag}\simeq \rma\equiv R\left(\frac{\bin}{2}\right)^{1/3}.
	\label{eq:rm}
\end{equation}
This expression is valid for $\bin\gg1$. When $\bin\lesssim 1$, $\rma$ becomes
less than $R$ and is no longer relevant for the system. Thus, the conductor's
effective radius, which we label as $\Reff$, is $R$ in the $\bin\ll1$ limit, and
proportional to $\rma$ in the $\bin\gg1$ limit. The exact connection of these
two limits depends on the orientation of $\vb{\Bin}$. In the anti-aligned case,
the effective radius is given by the location of the magnetic null points, where
$\Bbkg=0$ (see Eq. \ref{eq:Bbkg}). The system has dipole moment direction
$\vu{m}=-\vu{x}$, and the null points are confined to the $x$-axis. With these
constraints, we can solve for the radius satisfying $\Bbkg=0$, giving
\begin{equation}
  \Reff^{(\SIdeg{180})}
  =R\left[1+2\left(\flatfrac{\rma}{R}\right)^3\right]^{\flatfrac{1}{3}}
  =R\left(1+\bin\right)^{1/3},
\end{equation}
where the superscript ``$(\SIdeg{180})$'' implies that the equation applies to
the anti-aligned system. If we replace $R$ with $\Reff^{(\SIdeg{180})}$ in
Eq.~(\ref{eq:drag}), we find that the drag coefficient for the anti-aligned
magnetic configuration is
\begin{equation}
  \CA^{(\SIdeg{180})}=C_{A0}(\Reff^{(\SIdeg{180})} /R)^2
  =C_{A0}\left(1+\bin\right)^{2/3},
\label{eq:antidrag}
\end{equation}
where $C_{A0}\equiv \CA(\bin=0)\simeq 1.2$ for the adopted fiducial parameters.
The dashed curve in Figure \ref{fig:CAB} depicts Eq.~(\ref{eq:antidrag}), which
closely reproduces our numerical results for the anti-aligned system.

\figCAB

\figvcomp

More importantly, Figure \ref{fig:CAB} shows that deviation from anti-alignment
can significantly enhance the drag coefficient. For the aligned case
($\theta_m=0$) in the $\bin\gg1$ limit, our numerical result can be approximated
by
\begin{equation}
\CA^{(\SIdeg{0})}\simeq 20\,\bin^{2/3}\quad (\bin\gg 1). \label{eq:aldrag}
\end{equation}
In this limit, $\CA^{(\SIdeg{0})}$ follows the same $\bin^{2/3}\propto\rma^2$
scaling as Eq.~(\ref{eq:antidrag}), albeit with a much larger prefactor,
suggesting that the effective size of the magnetosphere also contributes to the
drag force in this case. Despite this agreement in scaling, even for $\bin\sim
1$ the difference in $\CA$ between the aligned and anti-aligned cases is about
an order of magnitude. We discuss the origin of this difference in Section
\ref{sec:analysis}. The enhancement over the anti-aligned case decreases with
$\thetam$, but it remains robust up to $\thetam\sim\SIdeg{120}$. Despite strong
dependence of $\CA$ on $\thetam$, we only observe modest changes in its value
when changing $\phim$.

\figanglecomp

\figfieldlinesurvey

\figfieldlineflux

In Figure \ref{fig:vcomp}, we show the effect of varying $\MA$ in the aligned
and anti-aligned cases for $\bin=5$. For both orientations, the drag coefficient
approaches a constant value (independent of $\MA$) when $\MA\lesssim 0.1$. The
flattening of the $\CA$--$\MA$ curve supports the validity of the \Alfven wing
model in the sub-\Alfvenic regime. We consider the transition to the
super-\Alfvenic regime in Section \ref{sec:supalfvenic}.

%%%%%%%%%%%%%%%%%%%%%%
\subsection{Analysis} \label{sec:analysis}
\subsubsection{Linked Flux}

Figure \ref{fig:anglecomp} highlights the strong dependence of the drag
coefficient on the relative orientation between $\vb{\Bin}$ and $\vb{\Bout}$.
How can we understand this dependence?

From Figure \ref{fig:fidvisual}, we see that the most important difference
between the aligned and anti-aligned cases is the magnetic flux (or the number
of field lines) connecting the conductor to infinity. Since the \Alfven wings
trace such field lines, it is natural to expect that the increase in this
``linked flux'' leads to the enhancement of $\CA$.

Figure \ref{fig:fieldlinesurvey} shows some examples of the $\vb{\Bbkg}$ field
lines for different $\theta_m$, with $\bin=100$ and $\bin=1$. We see that, in
general, there are three distinct regions: a closed field line region (around
the conductor), an ``outside'' region (where the field lines do not intercept
the conductor), and a ``linked'' region (where field lines from the conductor
extend to a large distance and merge with $\vb{B}_0$). The closed region is
confined approximately by the magnetic null points, where $\vb{\Bbkg}=0$.

To better understand the connection between $\CA$ and the linked field lines, we
numerically measure the flux of the conductor's linked field lines. Since the
flow is sub-Alfv\'{e}nic, we have $|\dB|\ll\Bbkg$, so the background field is a
suitable approximation to the field of the steady-state system. To perform the
measurement, we subdivide the surface of the conductor into a domain of surface
elements. From each surface element, we trace a $\vb{\Bbkg}$ field line, and if
the field line escapes to infinity, we count its corresponding surface element
as ``linked.'' We then integrate $\vb{\Bbkg}\vdot\vu{r}$ over all linked surface
elements to obtain their total flux, which we label $\PhiL$.

We show $\PhiL$ as a function of $\bin$ for several values of $\thetam$ in
Figure \ref{fig:fieldlineflux}(a). For all curves, we find two clear limits,
given by,
\begin{equation}
	\PhiL\propto
	\begin{cases}
		\pi R^2 \Bin,\quad \bin\lesssim1, \\
		\pi \rma^2 B_0,\quad \bin\gg1. \label{eq:philscale}
	\end{cases}
\end{equation}
These limits represent distinct physical pictures. In the $\bin\lesssim1$ limit,
which is depicted by the $\bin=1$ cases in Figure \ref{fig:fieldlinesurvey}
(bottom row), the background field dominates over the conductor's own dipole
field. As a result, even for misaligned cases (see the bottom-middle panel), all
the conductor's field lines open up. This means that $\PhiL$ is equal to the
conductor's total flux. We call this case ``saturated'' since all possible flux
is linked.

By contrast, in the $\bin\gg1$ limit, depicted by the $\bin=100$ cases in Figure
\ref{fig:fieldlinesurvey}, the dipole is strong enough to create a magnetosphere
that dominates over the background field. In this case, there are always some
closed field lines, so $\PhiL$ cannot keep up with the conductor's total flux.
Instead, we can consider how many field lines approaching from infinity reach
the conductor. Field lines are likely to link if they fall within the influence
of the conductor's magnetosphere. The linked flux, then, is roughly given by the
number of $\vb{B_0}$ field lines passing through the cross-section of the
magnetosphere, giving the $\bin\gg1$ limit of Eq.~(\ref{eq:philscale}).

To provide a fitting formula for $\PhiL$ at a given value of $\thetam$, we may
capture the above limits using the following expression,
\begin{equation}
  \PhiL(\thetam)=
\pi R^2 \Bin \left[1+\left(\frac{2\rma}{q R}\right)^p\right]^{-\flatfrac{1}{p}},
\label{eq:phi0ap}
\end{equation}
where $q(\thetam),~p(\thetam)$ are fitting parameters that control the location
and sharpness of the transition between the two regimes. These fits are shown in
Figure \ref{fig:fieldlineflux}(a).

In Figure \ref{fig:fieldlineflux}(b), we show $\PhiL$ as a function of $\thetam$
for several values of $\bin$. We normalize the curves using $\Phi_0$, the flux
value at $\thetam=\SIdeg{0}$. For $\bin\lesssim 1$, the curves remain flat for
low values of $\thetam$ due to the previously described saturation effect.
Beyond a critical value of $\thetam$, however, the dipole is twisted enough that
its field lines begin to close and $\PhiL$ decreases.

For $\bin\gtrsim 10$, we find that the fit
\begin{equation}
\PhiL(\thetam) \simeq \PhiL(0)\,\cos(\thetam/2),\quad (\bin\gtrsim 10)
\label{eq:highflux1}\end{equation}
where
\begin{equation}
\PhiL(0)\simeq 3 \pi\rma^2 B_0, 
  \label{eq:highbflux}
\end{equation}
describes the data very well [see Fig.~\ref{fig:fieldlineflux}(b)]. We are not
able to derive this specific trigonometric relationship
analytically.\footnote{OpenAI Codex derived an analytical expression, and
confirmed that it is given by $\cos(\thetam/2)$ multiplied by a complicated
function which varies by less than $\SI{6}{\percent}$ for all values of
$\theta_m$ (Tu Guo, private communication).} The $\cos(\thetam/2)$ dependence
was also suggested in previous works \citep{Saur+13,Presa+26}.

\subsubsection{Power}

To see how the linked magnetic field affects the \Alfven power, consider the
original picture of \Alfven wing \citep{Drell+65}. Suppose a ``linked'' magnetic
field $B_L$ connects a moving conductor with effective area ${\cal A}$ and the
background field. The \Alfven power is given by
\begin{equation}
  P\sim {1\over 4\pi}\left({B_Lv_0\over v_A}\right)^2 v_A  (2 {\cal A})
={v_0^2\over 2\pi}\sqrt{4\pi\rho_0} \left(B_L {\cal A}\right).
\end{equation}
In general, we replace $B_L{\cal A}$ by the linked flux and obtain
\begin{equation}
	P\sim \frac{v_0^2}{2\pi}\sqrt{4\pi\rho_0}\,\PhiL. \label{eq:genpower}
\end{equation}

For $\thetam\simeq 180^\circ$, although there is no linked flux in the
unperturbed field ${\bf B}_{\rm bkg}$ (see Fig.~\ref{fig:fieldlinesurvey}), the
conductor perturbs the background magnetic field to induce an effective linkage.
Thus $\PhiL$ may be replaced by an effective flux, $\pi(\Reff^{(\SIdeg{180})})^2
B_0$, consisting of the background field lines warped around a sphere of radius
$\Reff^{(\SIdeg{180})}$. The corresponding power can be written as
\begin{align}
  P (\thetam\simeq 180^\circ) &= \alpha_0 \frac{v_0^2}{2\pi}
  \sqrt{4\pi\rho_0}\,B_0
  \,\pi (\Reff^{(\SIdeg{180})})^2\nonumber\\
&=  2\pi \alpha_0 (1+\bin)^{2/3} \rho_0 v_0^2\vAn R^2, \label{eq:P0}
\end{align}
where $\alpha_0$ is a constant. This implies (see also Eq.~\ref{eq:antidrag})
\begin{align}
  \CA(\thetam\simeq 180^\circ) &=2\pi \alpha_0 (1+\bin)^{2/3}\nonumber\\
  & \simeq 1.2\,(1+\bin)^{2/3},
\label{eq:CA180}\end{align}
where the coefficient in the second line comes from our numerical simulation.

For the $\bin\gtrsim 10$ systems that are not strongly anti-aligned, combining
Eqs.~(\ref{eq:highflux1})--(\ref{eq:highbflux}) with Eq.~(\ref{eq:genpower}), we
have
\begin{align}
  P (\thetam) &= \alpha \frac{v_0^2}{2\pi}\sqrt{4\pi\rho_0}\,
\left(3 \pi\rma^2 B_0\right)\cos(\thetam/2) \nonumber\\
&= {6\pi\over 2^{2/3}}\alpha \bin^{2/3}\,
\rho_0 v_0^2\vAn R^2\cos(\thetam/2), \label{eq:power}
\end{align}
where $\alpha$ is a constant. This implies
\begin{align}
  \CA(\thetam) &={6\pi\over 2^{2/3}}\alpha \bin^{2/3}\cos(\thetam/2)\nonumber\\
  & \simeq 20\,\bin^{2/3}\cos(\thetam/2),\qquad
  (\bin\gtrsim 1,\,\thetam\neq 180^\circ)
\label{eq:CA}\end{align}  
where the coefficient in the second line comes from our numerical simulation
(see Eq.~\ref{eq:aldrag}).

\figfidv

Note that Eqs.~(\ref{eq:CA180}) and (\ref{eq:CA}) apply to different limiting
parameter regimes of $\bin$ and $\thetam$, and have different physical origins:
Eq.~(\ref{eq:CA180}) arises from ``induced'' magnetic linkage between the
conductor and the external magnetic field, while Eq.~(\ref{eq:CA}) arises from
``intrinsic'' magnetic linkage. Combining Eqs.~(\ref{eq:CA180}) and
(\ref{eq:CA}), we obtain a general expression for the drag coefficient that is
valid in various regimes:
\begin{equation}
  \CA(\bin,\thetam)\simeq 1.2\,(1+\bin)^{2/3}+ 20\,\bin^{2/3}\cos(\thetam/2).
  \label{eq:general}
\end{equation}
In particular, for $\bin\ll 1$, Eq.~(\ref{eq:general}) gives the correct $B_{\rm
in}=0$ limit ($\CA\simeq 1.2$) studied in Paper I; for $\bin\gtrsim 10$, it
recovers Eqs.~(\ref{eq:CA180}) and (\ref{eq:CA}). For $\bin \sim 1$, the
$\cos(\thetam/2)$ factor does not apply [see Fig.~\ref{fig:fieldlineflux}(b)],
but Eq.~(\ref{eq:general}) can still provide a reasonable approximation to the
drag coefficient.

%%%%%%%%%%%%%%%%%%%%%%%%
\section{Super-Alfv\'enic Regime} \label{sec:supalfvenic}

To this point, we have been considering the sub-\Alfvenic regime, where
$\MA\lesssim 1$. In this case, the power is largely dissipated from the system
in the form of \Alfvenic disturbances propagating along magnetic field lines and
can be described by Eq.~(\ref{eq:drag}), with the drag coefficient that depends
weakly on $\MA$ (see Eq.~\ref{eq:dragparam}) and can be approximated by
Eq.~(\ref{eq:general}). We now turn our attention to the super-\Alfvenic regime,
where $\MA\gtrsim 1$. This regime is commonly found in the context of
solar-system planets interacting with solar wind (see
\citealt{Kivelson&Russell95,Eastwood+15,Dong+19}). The parameters of our high
$\MA$ runs are already described in Section \ref{sec:subresults}.

In Figure \ref{fig:fidv}, we show the magnetic field configuration in the steady
state of our fiducial super-\Alfvenic run (aligned, $\bin=5$, $\MA=3$,
$\beta=2$). Rather than the \Alfven wing structure seen in the sub-\Alfvenic
runs, a series of shocks develop where the flow collides with the conductor. We
highlight the features relevant to our system below:
\begin{itemize}
\item The undisturbed upstream flow is decelerated and compressed at the
  \textit{bow shock}, which forms the outermost boundary of the interaction.
\item Downstream of the shock, the plasma and open magnetic field lines pile up
  in the \textit{magnetosheath}, where both thermal pressure and magnetic
  pressure are high, and the flow is deflected around the conductor.
\item On the upstream face, the accumulated magnetic pressure of the sheath
  balances the conductor's internal magnetic field at the magnetopause (a
  contact discontinuity), where the bulk flow stagnates, the magnetic field is
  strongly draped, and plasma flow is largely inhibited.
\item Behind the magnetopause, field lines linked to the conductor form two
  lobes with oppositely directed magnetic fields. These lobes extend far
  downstream in a structure called the \textit{magnetotail}. The lobes
  eventually become close enough that reconnection can occur between them, and a
  strong current sheet develops.
\item Many of the conductor's field lines are contained to a local magnetosphere
  between the lobes.
\end{itemize}
The above paradigm has been extensively studied, especially in the planetary
context, so we refer the reader to the literature for further discussion of the
above features (\textit{e.g.} \citealt{Eastwood+15}). For our purposes, their
identification serves both as a validation of our model's capacity to reproduce
known physics and as a qualitative demonstration that our model has reached a
new regime.

We quantitatively investigate this regime change through measurements of $\CA$,
which are shown in Figure \ref{fig:vcomp} (the sub-\Alfvenic regime was
discussed in Section \ref{sec:subresults}). In the $\MA\gtrsim1$ regime, the
splitting in $\CA$ between aligned and anti-aligned cases collapses. In the
aligned case, the drag force plummets, whereas it steadily increases in the
anti-aligned case. Beyond $\MA\sim$ a few, both trends converge to be
approximately linear in $\MA$ although we have not simulated $\MA>30$ cases in
this paper to systematically explore this trend.

We propose that, in the super-\Alfvenic regime, ram pressure overtakes the
\Alfven wing model as the dominant drag contribution. In other words, we may
consider the conventional drag coefficient, $\CD$, defined by the expression,
\begin{equation} 
	P = \CD \rho_0 \Reff^2 v_0^3, \label{eq:ramdrag}
\end{equation}
where we have used $\Reff$ as a stand-in for the conductor's effective radius,
accounting for the size of its magnetosphere (see Eq.~\ref{eq:rmag}). When the
drag force is dominated by ram pressure, we expect $\CD$ to be of order unity
and roughly constant. Equating Eq.~(\ref{eq:drag}) with Eq.~(\ref{eq:ramdrag})
gives
\begin{equation}
	\CA = \CD \MA (\Reff/R)^2. \label{eq:CD}
\end{equation}
This matches the trend observed in the $\MA\gg1$ limit in Figure
\ref{fig:vcomp}. For example, using the parameters of Figure \ref{fig:vcomp},
which set $\Reff\simeq\rma\simeq1.4R$, we find $\CD\simeq 1.0$. A detailed
exploration of how $\CD$ depends on various parameters in the super-\Alfvenic
regime is beyond the scope of this paper.

%%%%%%%%%%%%%%%%%%%%%%%
\section{Summary and Conclusions} \label{sec:discuss}

We have performed 3D MHD simulations of magnetized flow past a magnetized
spherical conductor and numerically measured the drag force on the conductor. We
introduce an \Alfvenic drag coefficient $\CA$ (see Eq.~1) to characterize the
problem. In Paper I \citep{Corso&Lai26}, we considered a non-magnetic conductor
and determined $\CA$ as a function of the \Alfven Mach number $\MA$ and the
plasma beta $\beta$ (see Eq.~2). Here, by introducing an internal magnetic field
${\bf B}_{\rm in}$ to the conductor, we have extended the model to include two
additional parameters: the relative strength of the magnetic field,
$\bin=\Binout$, and its orientation, defined through Eq.~(\ref{eq:moment}) by
the angles $\thetam$ and $\phim$. Alongside these new parameters, we have also
extended our model by probing super-\Alfvenic flows with $\MA\gtrsim1$.

Our study has revealed new features for the power dissipation and drag force in
different parameter regimes. Most importantly, we find that $\CA$ is greatly
enhanced from the $B_{\rm in}=0$ value when $\bin\gtrsim 1$ and $\thetam$ is not
too close to $180^\circ$ (see Fig. \ref{fig:CAB}). We show that this enhancement
is largely driven by open magnetic field lines that are linked to the conductor.
Power is efficiently dissipated by \Alfvenic disturbances propagating along
these field lines, which carry greater magnetic flux compared to the
anti-aligned case ($\thetam=180^\circ$) or non-magnetic case ($B_{\rm in}=0$).
We summarize these findings through the expressions for $\CA$ in different
parameter regimes (Eqs.~\ref{eq:CA180} and \ref{eq:CA}), and we provide an
approximate general expression for $\CA$ in Eq.~(\ref{eq:general}).

We have also explored the effect of varying $\MA$ on the drag coefficient. We
find that while the sub-\Alfvenic regime is characterized by \Alfven wings and
the related scaling laws for $\CA$, the super-\Alfvenic regime exhibits a
qualitatively different behavior. For $\MA\gtrsim 1$, the \Alfven wing picture
breaks down, and the drag force is instead dominated by ram pressure. We
summarize this effect through a new expression for $\CA$ in Eq.~(\ref{eq:CD}).

Our numerical results may have important ramifications for magnetic interactions
in binaries. For example, recent observations have revealed possible evidence
that a close-in planet can induce flares on its host star through magnetic
interactions \citep{Ilin+25,Revilla+26}; the observed flare power appears to
exceed the prediction of \Alfven wing models, and alternative models are often
invoked (\textit{e.g.} \citealt{Lanza09,Lanza13,Strugarek16}). However, most
\Alfven wing models to-date only consider the effective size of the planet's
magnetosphere when accounting for its internal magnetic field
(\textit{e.g.}~\citealt{Saur+13}), corresponding to our result for anti-aligned
systems (Eq.~\ref{eq:P0}). We have shown in this paper that the \Alfven wing
power may be enhanced by at least an order of magnitude when the planet's
magnetic field is more aligned with the surrounding field from its host star
(Eq.~\ref{eq:power}). This enhancement may be sufficient to explain the flare
power discrepancy and retain the viability of the \Alfven wing model.
Regardless, a proper understanding of the observational signatures of the
magnetic star-planet interactions requires a more sophisticated numerical
treatment of the plasma and radiative processes. This is our intended direction
for future work on such systems, and some efforts are already underway to this
end (\textit{e.g.}~\citealt{Lyutikov26}).

Concerning binary magnetic interactions in general, we note that the current
study is confined to the local region around the conductor, which is assumed to
be weakly magnetized compared to its primary companion. Our model cannot capture
more global effects that involve large-scale interactions in the system. As
such, the local physical processes studied in this paper may be occurring in
conjunction with some of the global processes (\textit{e.g.}~\citealt{Lai12,
  Strugarek16,Beloborodov21,Most&Philippov22,Most&Philippov23,
  Mahlmann&Beloborodov25}). Nevertheless, our finding of the enhanced drag/power
for the $\thetam\ne 180^\circ$ systems is consistent with previous global
studies of binary interactions. For example, \cite{Strugarek16} carried out
simulations of magnetic stellar wind interacting with a magnetic planet and
found that the aligned case exhibits much stronger interaction than the
anti-aligned case. For coalescing binary neutron stars,
\cite{Palenzuela+13a,Palenzuela+13b} and
\cite{Most&Philippov20,Most&Philippov22} carried out force-free electrodynamics
simulations for a range of magnetic field configurations. For binaries with
anti-aligned magnetic dipoles, which correspond to our ``aligned cases'', and
for more general mis-aligned dipoles, they found that a magnetic flux tube
connects the two stars and gets twisted by their rotation, leading to
quasi-periodic magnetic flares. In contrast, for binaries with aligned dipoles,
a very different dissipation mechanism operates, which relies on
Kelvin--Helmholtz-type instability at the magnetospheric interface, generating
Alfvénic turbulence and escaping fast magnetosonic waves
\citep{Mahlmann&Beloborodov25}.

Despite the generally broader coverage of the parameter space in this study
compared to Paper I, this coverage is not complete. For example, although Paper
I considered different values of $\beta$, we have held it fixed in this study
for simplicity. Our expectation is that the effect of varying $\beta$ would be
subdominant compared to the other parameters explored in this paper. Our model
also currently assumes that the plasma is ideal, which neglects the effects of
viscosity and resistivity. Their associated Reynolds numbers could serve as
additional dimensionless parameters of our system and introduce new dissipation
mechanisms not captured by our current model.

\section*{Acknowledgments}
This research was supported in part by the NSF grant DGE-2139899. Claude Code
was used to assist with code development in this research. We thank Nils Deppe,
Chuanfei Dong, Lawrence Kidder, Elias Most, Anatoly Spitkovsky and Jim Stone for
useful discussions.

\textit{Software:} Claude Code \citep{Anthropic25}, \texttt{Matplotlib}
\citep{Hunter07}, \texttt{NumPy} \citep{Harris+20}, \texttt{ParaView}
\citep{Ayachit15}.

\bibliography{Bibliography}{} \bibliographystyle{mnras}

\begin{thebibliography}{}
\makeatletter
\relax
\def\mn@urlcharsother{\let\do\@makeother \do\$\do\&\do\#\do\^\do\_\do\%\do\~}
\def\mn@doi{\begingroup\mn@urlcharsother \@ifnextchar [ {\mn@doi@}
  {\mn@doi@[]}}
\def\mn@doi@[#1]#2{\def\@tempa{#1}\ifx\@tempa\@empty \href
  {http://dx.doi.org/#2} {doi:#2}\else \href {http://dx.doi.org/#2} {#1}\fi
  \endgroup}
\def\mn@eprint#1#2{\mn@eprint@#1:#2::\@nil}
\def\mn@eprint@arXiv#1{\href {http://arxiv.org/abs/#1} {{\tt arXiv:#1}}}
\def\mn@eprint@dblp#1{\href {http://dblp.uni-trier.de/rec/bibtex/#1.xml}
  {dblp:#1}}
\def\mn@eprint@#1:#2:#3:#4\@nil{\def\@tempa {#1}\def\@tempb {#2}\def\@tempc
  {#3}\ifx \@tempc \@empty \let \@tempc \@tempb \let \@tempb \@tempa \fi \ifx
  \@tempb \@empty \def\@tempb {arXiv}\fi \@ifundefined
  {mn@eprint@\@tempb}{\@tempb:\@tempc}{\expandafter \expandafter \csname
  mn@eprint@\@tempb\endcsname \expandafter{\@tempc}}}

\bibitem[\protect\citeauthoryear{{Anthropic}}{{Anthropic}}{2025}]{Anthropic25}
{Anthropic} 2025, Claude Code,
  \url{https://docs.anthropic.com/en/docs/claude-code/overview}

\bibitem[\protect\citeauthoryear{Ayachit}{Ayachit}{2015}]{Ayachit15}
Ayachit U.,  2015, The ParaView Guide: A Parallel Visualization Application.
Kitware, Inc., Clifton Park, NY, USA

\bibitem[\protect\citeauthoryear{Bagenal \& Dols}{Bagenal \&
  Dols}{2020}]{Bagenal&Dols20}
Bagenal F.,  Dols V.,  2020, \mn@doi [J. Geophys. Res. Space Phys.]
  {10.1029/2019JA027485}, 125, e2019JA027485

\bibitem[\protect\citeauthoryear{Beloborodov}{Beloborodov}{2021}]{Beloborodov21}
Beloborodov A.~M.,  2021, \mn@doi [\apj] {10.3847/1538-4357/ac17e7}, 921, 92

\bibitem[\protect\citeauthoryear{{Bigg}}{{Bigg}}{1964}]{Bigg64}
{Bigg} E.~K.,  1964, \mn@doi [\nat] {10.1038/2031008a0}, \href
  {https://ui.adsabs.harvard.edu/abs/1964Natur.203.1008B} {203, 1008}

\bibitem[\protect\citeauthoryear{{Corso} \& {Lai}}{{Corso} \&
  {Lai}}{2026}]{Corso&Lai26}
{Corso} N.~J.,  {Lai} D.,  2026, \mn@doi [\apj] {10.3847/1538-4357/ae5817},
  \href {https://ui.adsabs.harvard.edu/abs/2026ApJ..1002....1C} {1002, 1}

\bibitem[\protect\citeauthoryear{De~Colle, Lin, Chen  \& Li}{De~Colle
  et~al.}{2025}]{DeColle+25}
De~Colle F.,  Lin D. N.~C.,  Chen C.,   Li G.,  2025, \mn@doi [\mnras]
  {10.1093/mnras/staf707}, 539, 3459

\bibitem[\protect\citeauthoryear{{Dong}, {Wang}, {Hakim}, {Bhattacharjee},
  {Slavin}, {DiBraccio}  \& {Germaschewski}}{{Dong} et~al.}{2019}]{Dong+19}
{Dong} C.,  {Wang} L.,  {Hakim} A.,  {Bhattacharjee} A.,  {Slavin} J.~A.,
  {DiBraccio} G.~A.,   {Germaschewski} K.,  2019, \mn@doi [\grl]
  {10.1029/2019GL083180}, \href
  {https://ui.adsabs.harvard.edu/abs/2019GeoRL..4611584D} {46, 11,584}

\bibitem[\protect\citeauthoryear{Drell, Foley  \& Ruderman}{Drell
  et~al.}{1965}]{Drell+65}
Drell S.~D.,  Foley H.~M.,   Ruderman M.~A.,  1965, \mn@doi [\prl]
  {10.1103/PhysRevLett.14.171}, 14, 171

\bibitem[\protect\citeauthoryear{{Eastwood}, {Hietala}, {Toth}, {Phan}  \&
  {Fujimoto}}{{Eastwood} et~al.}{2015}]{Eastwood+15}
{Eastwood} J.~P.,  {Hietala} H.,  {Toth} G.,  {Phan} T.~D.,   {Fujimoto} M.,
  2015, \mn@doi [\ssr] {10.1007/s11214-014-0050-x}, \href
  {https://ui.adsabs.harvard.edu/abs/2015SSRv..188..251E} {188, 251}

\bibitem[\protect\citeauthoryear{{Goldreich} \& {Lynden-Bell}}{{Goldreich} \&
  {Lynden-Bell}}{1969}]{Goldreich&Lynden-Bell69}
{Goldreich} P.,  {Lynden-Bell} D.,  1969, \mn@doi [\apj] {10.1086/149947},
  \href {https://ui.adsabs.harvard.edu/abs/1969ApJ...156...59G} {156, 59}

\bibitem[\protect\citeauthoryear{Hansen \& Lyutikov}{Hansen \&
  Lyutikov}{2001}]{Hansen&Lyutikov01}
Hansen B. M.~S.,  Lyutikov M.,  2001, \mn@doi [\mnras]
  {10.1046/j.1365-8711.2001.04103.x}, 322, 695

\bibitem[\protect\citeauthoryear{Harris et~al.,}{Harris
  et~al.}{2020}]{Harris+20}
Harris C.~R.,  et~al., 2020, \mn@doi [Nature] {10.1038/s41586-020-2649-2}, 585,
  357

\bibitem[\protect\citeauthoryear{Hunter}{Hunter}{2007}]{Hunter07}
Hunter J.~D.,  2007, \mn@doi [Comput. Sci. Eng.] {10.1109/MCSE.2007.55}, 9, 90

\bibitem[\protect\citeauthoryear{{Ilin}, {Vedantham}, {Poppenh{\"a}ger},
  {Bloot}, {Callingham}, {Brandeker}  \& {Chakraborty}}{{Ilin}
  et~al.}{2025}]{Ilin+25}
{Ilin} E.,  {Vedantham} H.~K.,  {Poppenh{\"a}ger} K.,  {Bloot} S.,
  {Callingham} J.~R.,  {Brandeker} A.,   {Chakraborty} H.,  2025, \mn@doi
  [\nat] {10.1038/s41586-025-09236-z}, \href
  {https://ui.adsabs.harvard.edu/abs/2025Natur.643..645I} {643, 645}

\bibitem[\protect\citeauthoryear{{Kivelson} \& {Russell}}{{Kivelson} \&
  {Russell}}{1995}]{Kivelson&Russell95}
{Kivelson} M.~G.,  {Russell} C.~T.,  eds, 1995, {Introduction to Space
  Physics}.
Cambridge University Press, Cambridge, UK, \mn@doi{10.1017/9781139878296}

\bibitem[\protect\citeauthoryear{Lai}{Lai}{2012}]{Lai12}
Lai D.,  2012, \mn@doi [\apjl] {10.1088/2041-8205/757/1/L3}, 757, L3

\bibitem[\protect\citeauthoryear{Laine \& Lin}{Laine \&
  Lin}{2011}]{Laine&Lin11}
Laine R.~O.,  Lin D. N.~C.,  2011, \mn@doi [\apj] {10.1088/0004-637X/745/1/2},
  745, 2

\bibitem[\protect\citeauthoryear{{Lanza}}{{Lanza}}{2009}]{Lanza09}
{Lanza} A.~F.,  2009, \mn@doi [\aap] {10.1051/0004-6361/200912367}, \href
  {https://ui.adsabs.harvard.edu/abs/2009A&A...505..339L} {505, 339}

\bibitem[\protect\citeauthoryear{{Lanza}}{{Lanza}}{2013}]{Lanza13}
{Lanza} A.~F.,  2013, \mn@doi [\aap] {10.1051/0004-6361/201321790}, 557, A31

\bibitem[\protect\citeauthoryear{{Lee} \& {Owen}}{{Lee} \&
  {Owen}}{2025}]{Lee&Owen25}
{Lee} E.~J.,  {Owen} J.~E.,  2025, \mn@doi [\apjl] {10.3847/2041-8213/adafa3},
  \href {https://ui.adsabs.harvard.edu/abs/2025ApJ...980L..40L} {980, L40}

\bibitem[\protect\citeauthoryear{Li, Ferrario  \& Wickramasinghe}{Li
  et~al.}{1998}]{Li+98}
Li J.,  Ferrario L.,   Wickramasinghe D.,  1998, \mn@doi [\apj]
  {10.1086/311546}, 503, L151

\bibitem[\protect\citeauthoryear{{Lyutikov}}{{Lyutikov}}{2026}]{Lyutikov26}
{Lyutikov} M.,  2026, \mn@doi [\apjl] {10.3847/2041-8213/ae7333}, \href
  {https://ui.adsabs.harvard.edu/abs/2026ApJ..1004L..43L} {1004, L43}

\bibitem[\protect\citeauthoryear{Mahlmann \& Beloborodov}{Mahlmann \&
  Beloborodov}{2025}]{Mahlmann&Beloborodov25}
Mahlmann J.~F.,  Beloborodov A.~M.,  2025, \mn@doi [\apjl]
  {10.3847/2041-8213/adb5fd}, 981, L17

\bibitem[\protect\citeauthoryear{Mignone, Bodo, Massaglia, Matsakos, Tesileanu,
  Zanni  \& Ferrari}{Mignone et~al.}{2007}]{Mignone07}
Mignone A.,  Bodo G.,  Massaglia S.,  Matsakos T.,  Tesileanu O.,  Zanni C.,
  Ferrari A.,  2007, \mn@doi [\apjs] {10.1086/513316}, 170, 228

\bibitem[\protect\citeauthoryear{Most \& Philippov}{Most \&
  Philippov}{2020}]{Most&Philippov20}
Most E.~R.,  Philippov A.~A.,  2020, \mn@doi [\apjl]
  {10.3847/2041-8213/ab8196}, \href
  {https://ui.adsabs.harvard.edu/abs/2020ApJ...893L...6M} {893, L6}

\bibitem[\protect\citeauthoryear{Most \& Philippov}{Most \&
  Philippov}{2022}]{Most&Philippov22}
Most E.~R.,  Philippov A.~A.,  2022, \mn@doi [\mnras] {10.1093/mnras/stac1909},
  \href {https://ui.adsabs.harvard.edu/abs/2022MNRAS.515.2710M} {515, 2710}

\bibitem[\protect\citeauthoryear{Most \& Philippov}{Most \&
  Philippov}{2023}]{Most&Philippov23}
Most E.~R.,  Philippov A.~A.,  2023, \mn@doi [\apjl]
  {10.3847/2041-8213/acfdae}, 956, L33

\bibitem[\protect\citeauthoryear{Neubauer}{Neubauer}{1980}]{Neubauer80}
Neubauer F.~M.,  1980, \mn@doi [J. Geophys. Res. Space Phys.]
  {10.1029/JA085iA03p01171}, 85, 1171

\bibitem[\protect\citeauthoryear{Neubauer}{Neubauer}{1990}]{Neubauer90}
Neubauer F.~M.,  1990, \mn@doi [Adv. Space Res.]
  {10.1016/0273-1177(90)90083-C}, 10, 25

\bibitem[\protect\citeauthoryear{Neubauer}{Neubauer}{1998}]{Neubauer98}
Neubauer F.~M.,  1998, \mn@doi [J. Geophys. Res. Planets] {10.1029/97JE03370},
  103, 19843

\bibitem[\protect\citeauthoryear{{Palenzuela}, {Lehner}, {Liebling}, {Ponce},
  {Anderson}, {Neilsen}  \& {Motl}}{{Palenzuela}
  et~al.}{2013a}]{Palenzuela+13a}
{Palenzuela} C.,  {Lehner} L.,  {Liebling} S.~L.,  {Ponce} M.,  {Anderson} M.,
  {Neilsen} D.,   {Motl} P.,  2013a, \mn@doi [\prd]
  {10.1103/PhysRevD.88.043011}, \href
  {https://ui.adsabs.harvard.edu/abs/2013PhRvD..88d3011P} {88, 043011}

\bibitem[\protect\citeauthoryear{{Palenzuela}, {Lehner}, {Ponce}, {Liebling},
  {Anderson}, {Neilsen}  \& {Motl}}{{Palenzuela}
  et~al.}{2013b}]{Palenzuela+13b}
{Palenzuela} C.,  {Lehner} L.,  {Ponce} M.,  {Liebling} S.~L.,  {Anderson} M.,
  {Neilsen} D.,   {Motl} P.,  2013b, \mn@doi [\prl]
  {10.1103/PhysRevLett.111.061105}, \href
  {https://ui.adsabs.harvard.edu/abs/2013PhRvL.111f1105P} {111, 061105}

\bibitem[\protect\citeauthoryear{{Presa}, {Vidotto}  \& {Elekes}}{{Presa}
  et~al.}{2026}]{Presa+26}
{Presa} A.,  {Vidotto} A.~A.,   {Elekes} F.,  2026, \mn@doi [\mnras]
  {10.1093/mnras/stag1081}, \href
  {https://ui.adsabs.harvard.edu/abs/2026MNRAS.549g1081P} {549, stag1081}

\bibitem[\protect\citeauthoryear{Preusse, Kopp, Büchner  \&
  Motschmann}{Preusse et~al.}{2006}]{Preusse+06}
Preusse S.,  Kopp A.,  Büchner J.,   Motschmann U.,  2006, \mn@doi [\aap]
  {10.1051/0004-6361:20065353}, 460, 317

\bibitem[\protect\citeauthoryear{{Revilla} et~al.,}{{Revilla}
  et~al.}{2026}]{Revilla+26}
{Revilla} D.,  et~al., 2026, \mn@doi [Science] {10.1126/science.adv3075}, \href
  {https://ui.adsabs.harvard.edu/abs/2026Sci...393..403R} {393, 403}

\bibitem[\protect\citeauthoryear{{Saur}}{{Saur}}{2018}]{Saur18}
{Saur} J.,  2018, in {Deeg} H.~J.,  {Belmonte} J.~A.,  eds, Handbook of
  Exoplanets.
Springer International Publishing, Cham, pp 1877--1893,
  \mn@doi{10.1007/978-3-319-55333-7_27}

\bibitem[\protect\citeauthoryear{{Saur}, {Grambusch}, {Duling}, {Neubauer}  \&
  {Simon}}{{Saur} et~al.}{2013}]{Saur+13}
{Saur} J.,  {Grambusch} T.,  {Duling} S.,  {Neubauer} F.~M.,   {Simon} S.,
  2013, \mn@doi [\aap] {10.1051/0004-6361/201118179}, \href
  {https://ui.adsabs.harvard.edu/abs/2013A&A...552A.119S} {552, A119}

\bibitem[\protect\citeauthoryear{{Sharma}, {Turyshev}, {Barkov}  \&
  {Lyutikov}}{{Sharma} et~al.}{2026}]{Sharma+26}
{Sharma} P.,  {Turyshev} S.~G.,  {Barkov} M.~V.,   {Lyutikov} M.,  2026,
  \mn@doi [\apj] {10.3847/1538-4357/ae7c6e}, \href
  {https://ui.adsabs.harvard.edu/abs/2026ApJ..1006...45S} {1006, 45}

\bibitem[\protect\citeauthoryear{{Shu}}{{Shu}}{1992}]{Shu92}
{Shu} F.~H.,  1992, {The Physics of Astrophysics. Volume II: Gas Dynamics}.
University Science Books, Mill Valley, CA

\bibitem[\protect\citeauthoryear{{Strugarek}}{{Strugarek}}{2016}]{Strugarek16}
{Strugarek} A.,  2016, \mn@doi [\apj] {10.3847/1538-4357/833/2/140}, \href
  {https://ui.adsabs.harvard.edu/abs/2016ApJ...833..140S} {833, 140}

\bibitem[\protect\citeauthoryear{Strugarek, Brun, Matt  \& Réville}{Strugarek
  et~al.}{2014}]{Strugarek+14}
Strugarek A.,  Brun A.~S.,  Matt S.~P.,   Réville V.,  2014, \mn@doi [\apj]
  {10.1088/0004-637X/795/1/86}, 795, 86

\bibitem[\protect\citeauthoryear{Wang, Yang, Wu, Dai  \& Wang}{Wang
  et~al.}{2016}]{Wang+16}
Wang J.-S.,  Yang Y.-P.,  Wu X.-F.,  Dai Z.-G.,   Wang F.-Y.,  2016, \mn@doi
  [\apjl] {10.3847/2041-8205/822/1/L7}, 822, L7

\makeatother
\end{thebibliography}

\end{document}